# To slow, or not to slow?  New science in sub-second networks

Neil F. Johnson[1]

[1] Department of Physics, University of Miami, Coral Gables, FL 33126, USA

**Abstract**
**What happens when you slow down part of an ultrafast network that is operating quicker than the blink of an eye, e.g. electronic exchange network, navigational systems in driverless vehicles, or even neuronal processes in the brain? This question just adopted immediate commercial, legal and political importance following U.S. financial regulators' decision to allow a new network node to intentionally introduce delays of 350 microseconds. Though similar requests are set to follow, there is still no scientific understanding available to policymakers of the likely system-wide impact of such delays. Giving academic researchers access to (so far prohibitively expensive) microsecond exchange data would help rectify this situation. As a by-product, the lessons learned would deepen understanding of instabilities across myriad other networks, e.g. impact of millisecond delays on brain function and safety of driverless vehicle navigation systems beyond human response times.**

Fall 2016 brought a fundamental change to the United States. Its fastest and largest network – the decentralized network of electronic market exchanges – began to experience its first ever *intentional* delay. Specifically, a 38 mile coil of fiber-optic cable was embedded into a new exchange network node which, given the finite speed of light ($< 3.0 \times 10^8$ m/s in any material), introduced a systematic 350 microsecond delay *(1)*.

Nobody knows what future impacts this might have at the systems level.

The 'common good' justification used by policy makers is entirely reasonable, i.e. to level out highly asymmetric advantages available to faster participants. 350 microseconds sounds tiny (i.e. 0.35 milliseconds, or $3.5 \times 10^{-4}$ seconds) and is more than 1 million times quicker than the duration of the infamous 2010 'Flash' Crash ($\sim 10^3$ seconds) *(2)*. However today's electronic exchanges are an all-machine playing field with extreme sub-second operating times that lie way beyond the $\sim 1$ second real-time response and intervention of any human *(3,4)*. High-speed algorithms now receive, process, and respond to information on the scale of microseconds ($10^{-6}$ seconds) with the only guaranteed future speed barrier being the speed of light. Hundreds of orders are executed across multiple exchange nodes within 1 millisecond. The competitive need for speed is so great that new cables are being laid under oceans and through mountains to shave milliseconds off information transfer times *(5)*; new transnational networks of microwave communication towers are being constructed; new hard-wired semiconductor technology is being invented to reduce machine decision-making times down toward nanoseconds *(6)*; and new buildings are being purchased so that servers are geographically, and hence temporally, closer to other network nodes. So is 350 microseconds too much? Or too little? Should other exchange nodes be allowed to introduce their own delays? If so, how big? Might it even be possible that judicious policy management of such systematic delays could be used to *enhance* system-level stability and safety in ultrafast network systems?

The need to develop a systems-level understanding concerning such regulation in sub-second networks, is beginning to extend beyond the financial world. Two recent examples concern the navigational networks in driverless cars such as Uber that are appearing on the streets of Pittsburgh *(7)*, and unmanned aerial vehicles (drones) whose widespread commercial use was given the green light by the White House in August 2016 *(8)*. Given that the navigational processing in their underlying networks of sensors and software operates much faster than human response times, what regulatory principles should be hardwired or encoded in individual vehicles – or fleets of vehicles – in order to safely manage any systematic delays that arise from hardware or software defects *(9)*?

These diverse policy decisions will continue to be challenging and contentious until the core scientific question is fully addressed: What types of extreme behaviors can a given policy (i.e. perturbation such as an intentional time delay) generate in a decentralized, sub-second network of decision-making machinery where each component feeds off of information that is updated in time as new activity occurs? It is no accident that this question is also a primary one for the most complex network system of them all -- the human brain *(10,11)*. For example, it has been speculated that deep connections might exist between sub-second signal delays and extreme behavior.

The good news is that the electronic exchange system offers a unique opportunity to address this question, and in so doing benefit not only financial regulators but also policymakers and scientists in other domains. As well as being the largest and fastest societal network, it is the most data-rich, with every two-body interaction and one-body action (trades and quotes respectively) being recorded down to microsecond timescales. The problem is that as the timescale resolution of the data gets smaller, the cost of this data becomes prohibitively expensive. With a few exceptions *(*e.g. refs. 12-15*)*, this has forced academic studies to focus on price data on the timescale of minutes, hours, days, weeks and months. On these longer timescales, real-time human decision-making enters the dynamics and helps generate highly stochastic price behavior that eventually tends toward an unsurprising random walk. However the sub-second scale is all-machine, meaning that the system output can be strongly deterministic. This means that the system's output can move quickly in definite directions and hence generate extreme sub-second behaviors.

Figure 1 illustrates this point. It shows an example of extreme behavior associated with systematic delays that recently emerged within a time window of 500 milliseconds beyond any real-time human intervention. An association can be seen between the system-wide reporting of three distinct bunches of delayed signals (prices) shown by dotted diagonal arrows, and the onset of three new price features in Fig. 1A shown by solid vertical arrows. As each bunch is reported, automated trading algorithms (i.e. agents) that only have access to this slower information are given a false impression of a sudden increase in market activity. They quickly generate additional supply and demand, which then manifests itself a few milliseconds later as a new dynamical feature in the price of new trade events appearing across the network (Fig. 1A). Not only does this unexpected shockwave pattern of delays warn policy makers that they cannot rely on delays being independent in sub-second complex networks, it also suggests that extreme behaviors might be generated through a similar bunching of delays in other complex system domains. For example, it is intriguing that the millisecond timescale over which the system-wide extreme behavior emerges in Fig. 1A coincides with human cognitive processes in the brain, crossing from neuronal events to the emergence of human perception and consciousness *(10)*. An important takeaway message for both policy-makers and researchers is that there are likely to be different classes of extreme sub-second system behaviors classified by their sensitivity to such systematic delays. Fully developing such a classification could help policymakers tailor different versions of policies to cope with different market scenarios.

**References**

1. K. McCoy. IEX exchange readies for mainstream debut. August 18, 2016. Available at http://tabbforum.com/news/iex-exchange-readies-for-mainstream-debut?utm_campaign=d19dc21bc0-UA-12160392-1&utm_medium=email&utm_source=TabbFORUM%20Alerts&utm_term=0_29f4b8f8f1-d19dc21bc0-278278873. Accessed 08-18-2016
2. U.S. Commodity Futures Trading Commission. *Findings regarding the market events of May 6, 2010.* Available at http://www.sec.gov/news/studies/2010/marketevents-report.pdf
3. T.N. Liukkonen, K. Unit. Human reaction times as a response to delays in control systems – Notes in vehicular context (2009). Available at http://www.measurepolis.fi/alma/ALMA%20Human%20Reaction%20Times%20as%20a%20Response%20to%20Delays%20in%20Control%20Systems.pdf. Last visited: 17-2-2013
4. P. Saariluoma, Chess players' thinking: a cognitive psychological approach (Routledge, New York, 1995) p.43.
5. J. Pappalardo. New Transatlantic Cable Built to Shave 5 Milliseconds off Stock Trades. Available at http://www.popularmechanics.com/technology/engineering/infrastructure/a-transatlantic-cable-to-shave-5-milliseconds-off-stock-trades
6. B. Conway. Wall Street's Need For Trading Speed: The Nanosecond Age. Available at http://blogs.wsj.com/marketbeat/2011/06/14/wall-streets-need-for-trading-speed-the-nanosecond-age/
7. Uber to introduce self-driving cars within weeks. Available at http://www.bbc.com/news/technology-37117831. Accessed 08-18-2016
8. D. Kramer. White House encourages adoption of drones. *Physics Today*, August 9, 2016. Available at http://scitation.aip.org/content/aip/magazine/physicstoday/news/10.1063/PT.5.1080?utm_source=Physics%20Today&utm_medium=email&utm_campaign=7427978_The%20week%20in%20Physics%208–12%20August&dm_i=1Y69,4F7GQ,E1O09U,GA27B,1
9. A. Vespignani, Predicting the behavior of techno-social systems. *Science* **325**, 425 (2009).
10. D. Eagleman. How does the timing of neural signals map onto the timing of perception? in *Space and Time in Perception and Action*, Ed. Romi Nijhawan (Cambridge University Press, Cambridge, 2010)
11. N. Kasthuri, J.W. Lichtman. The role of neuronal identity in synaptic competition. *Nature* **424**, 426 (2003)
12. T. Preis, J.J. Schneider, H.E. Stanley, Switching Processes in Financial Markets. *Proceedings of the National Academy of Sciences* **108**, 7674 (2011).
13. J. Cartlidge, C. Szostek, M. De Luca and D. Cliff . "Too fast, too furious" In: 4th Int. Conf. Agents & Art. Intell 2, 126–135 (2012). Available at http://www.cs.bris.ac.uk/~cszjpc/docs/cartlidge-icaart-2012.pdf
14. R.N. Mantegna, H.E. Stanley, Scaling Behavior in the Dynamics of an Economic Index. *Nature* **376**, 46 (1995).
15. N.F. Johnson, G. Zhao, E. Hunsader, H. Qi, N. Johnson, J. Meng, B. Tivnan. Abrupt rise of new machine ecology beyond human response time. *Scientific Reports* 3, 2627 (2013). Available at http://www.nature.com/articles/srep02627

**Acknowledgments:** N.F.J. acknowledges funding from the National Science Foundation grant CNS1522693 and Air Force (AFOSR) grant FA9550-16-1-0247, and thanks Eric Hunsader and Pedro Manrique for their assistance with the data in Figure 1, and Brian Tivnan for discussions.

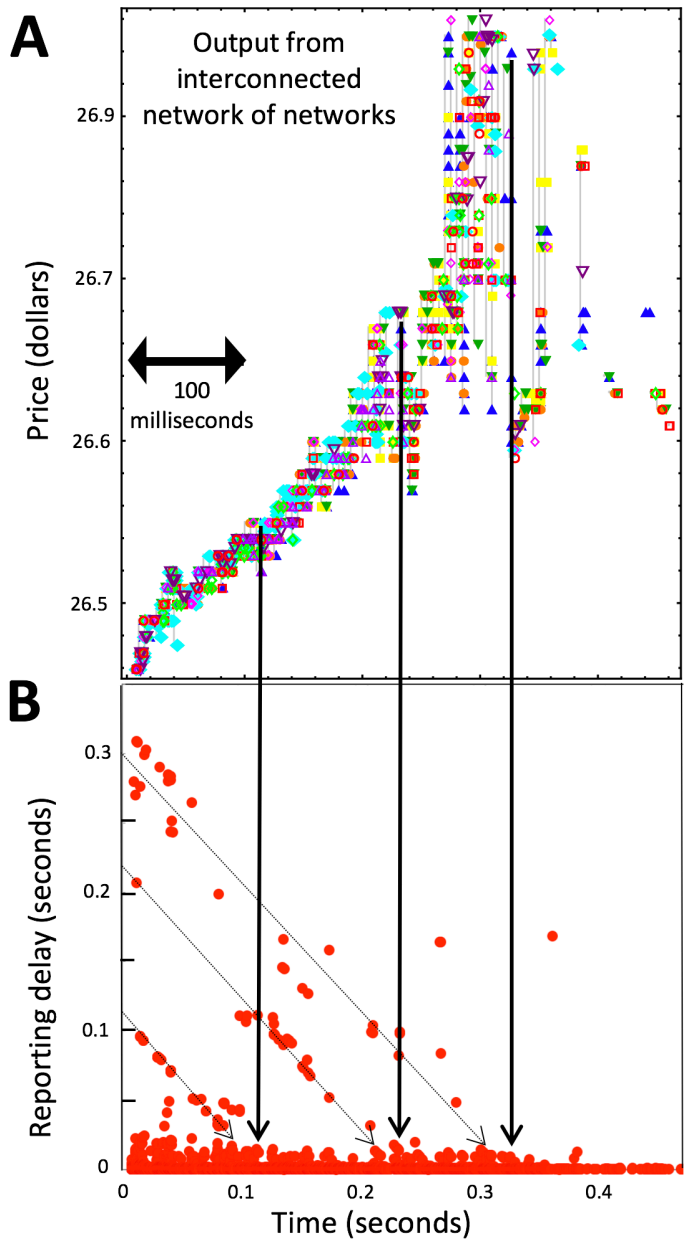

**Figure 1. A**: Extreme sub-second behavior in the network of networks of electronic exchanges at the heart of the financial system. Datapoints are individual events (i.e. trades) showing price of each trade on vertical scale, for Suncor stock during a half-second (i.e. 500 millisecond) time window ending at 14:39:02 on an otherwise typical day. The various electronic exchange networks, whose output is shown by separate colors, are physically connected to each other by communications channels, creating a network of networks across which information is shared. Gray vertical lines connect the lowest and highest price values occurring anywhere across the system at that instant: the larger the value, the larger the mismatch in consensus across the system. Data kindly provided by NANEX and extracted from the electronic exchange networks in the dominant financial region within 100 miles of Wall Street. **B**: Each red dot shows the delay (vertical scale) between the time at which a particular trade event and hence price is generated in a particular network exchange (which is shown on the horizontal scale) and the time at which it was reported system-wide. Though these delays predate the launch of the intentional 350 microsecond delay *(1)*, their strong temporal correlation serves to illustrate the impact that such intentional (and hence highly temporally correlated) delays can have. The dotted black diagonal arrows show the waves of successive trade events that get reported system-wide at the same time, indicated by solid black vertical arrows. These times are a few milliseconds ahead of significant dynamical features in Fig. 1A, i.e. onset of a sudden change in slope (left vertical line), onset of a sudden large fluctuation (middle vertical line) and the onset of a sudden collapse (right vertical line).